\documentclass{bioinfo}
\copyrightyear{2015}
\pubyear{2015}

\usepackage[hyphens]{url}
\usepackage{listings}
\usepackage{graphicx}
\usepackage{xspace}
\usepackage{amsmath}
\usepackage{authblk}
\usepackage{mathtools}
\usepackage{amsthm}
\usepackage{subfig}
\usepackage[countmax]{subfloat}
\usepackage{placeins}
\usepackage[outdir=./figures/]{epstopdf}

\newcommand{\vocab}[1]{\textbf{#1}\xspace}

\author[1]{Novak, Adam M}
\author[2]{Rosen, Yohei}
\author[1]{Haussler, David}
\author[1]{Paten, Benedict}
\affil[1]{UC Santa Cruz Genomics Institute, 1156 High Street, Santa Cruz, CA 95064}
\affil[2]{NYU School of Medicine, 550 First Avenue, New York, NY 10016}
\address{}
\title{Canonical, Stable, General Mapping using Context Schemes}
\history{Received on XXXXX; revised on XXXXX; accepted on XXXXX}
\editor{Associate Editor: XXXXXXX}

\begin{document}
\firstpage{1}

\setlength{\textfloatsep}{1pt plus 1.0pt}

\newtheorem{theorem}{Theorem}
\newtheorem{lemma}{Lemma}

\maketitle

\begin{abstract}
\section{Motivation:}
Sequence mapping is the cornerstone of modern genomics. However, most existing sequence mapping algorithms are insufficiently general.

\section{Results:}
We introduce context schemes: a method that allows the unambiguous recognition of a reference base in a query sequence by testing the query for substrings from an algorithmically defined set. Context schemes only map when there is a unique best mapping, and define this criterion uniformly for all reference bases. Mappings under context schemes can also be made stable, so that extension of the query string (e.g. by increasing read length) will not alter the mapping of previously mapped positions. Context schemes are general in several senses. They natively support the detection of arbitrary complex, novel rearrangements relative to the reference. They can scale over orders of magnitude in query sequence length. Finally, they are trivially extensible to more complex reference structures, such as graphs, that incorporate additional variation.
We demonstrate empirically the existence of high performance context schemes, and present efficient context scheme mapping algorithms.

\section{Availability and Implementation:}
The software test framework created for this work is available from \url{https://registry.hub.docker.com/u/adamnovak/sequence-graphs/}.

\section{Contact:} \href{benedict@soe.ucsc.edu}{benedict@soe.ucsc.edu}
\section{Supplementary Information:} Six supplementary figures and one supplementary section are available with the online version of this article.
\end{abstract}

\section{Introduction}


Many tools and algorithms exist for mapping reads to a reference genome \citep{li2010fast,langmead2009ultrafast,harris2007improved}. These tools are based on the idea of scoring local alignments between a query string and a reference according to some set of match, mismatch, and gap scoring parameters, and then finding local alignments with maximal or near-maximal scores. Seed-and-extend approaches coupled with memory efficient substring indexes or hashing schemes have been highly successful in heuristically accelerating this search process \citep{dobin2013star,li2010fast,langmead2009ultrafast}. 

The core problem with read mapping is ambiguity. There is often no single best place that a read maps, especially in the case of recent duplication within the reference genome. The precise base-level alignment of the read to a given location in the reference is also often ambiguous. To mitigate this, each mapped read is given a mapping quality, a per read score that indicates how likely the mapping was generated erroneously \citep{li2008mapping}. Quantifying this uncertainty is a reasonable approach for many applications, but even then the uncertainty can be difficult to accommodate downstream.

The difficulty of mapping a read to a reference motivates a consideration of its necessity. Recently, alignment-free methods of variant calling through substring detection have garnered significant interest \citep{dilthey2014improved}. The basic idea is not new; the dbSNP database has long provided, for each point variant in the database, a flanking nucleotide string that indicates the DNA context in which the variation was isolated \citep{sherry2001dbsnp}. In principle such a system of variant identification sidesteps the limitations of score based alignment, and can be used to canonically detect variations. However, in practice, insufficient rigor in defining the substrings to detect, and a failure to account for other variation near identified point mutations have limited the approach's usefulness. Here we formalize and extend this core idea; we propose using multiple, algorithmically defined context strings to canonically identify the presence of each base within a reference genome (potentially paving the way for high-specificity, alignment-free variant calling) and evaluate the performance of such a method in practice.

\section{Methods}






Throughout we make use of \vocab{DNA strings}, which are finite strings over the alphabet of \vocab{DNA bases} $\left\{\mathrm{A}, \mathrm{C}, \mathrm{G}, \mathrm{T}\right\}$. 
A DNA string $x$ has a \vocab{reverse complement} $x^*$, which is the reverse of $x$ with each DNA base replaced with its complement; $\mathrm{A}$ and $\mathrm{T}$ are complements of each other, as are $\mathrm{G}$ and $\mathrm{C}$. 

\subsection{Mapping}
A \vocab{reference (genome)} $G$ is a set of DNA strings and an index set of the elements of these strings, each member of which is called a \vocab{position}. Each position $p$ uniquely identifies an element $b(p)$ of a string in $G$. This allows us to unambiguously discuss the ``positions'' in that set of DNA strings, rather than ``bases'' or ``characters'', which could be interpreted as the four DNA bases themselves.

 We define the problem of mapping a \vocab{query} DNA string $x=(x_i)_{i=1}^n$ to a reference $G$. 
A \vocab{mapping scheme} is a function that takes $x$ and $G$ and, for each query element $i$ of $x$, either returns a position in $G$, declaring the query element $i$ to be \vocab{mapped} to that position in $G$, or declares the query element to be \vocab{unmapped} in $G$. For the scheme to map a query element to a position $p$ in $G$, $b(p)$ must either be $x_i$ (in which case that query element is \vocab{forward mapped}), or ${x_i}^*$ (in which case that query element is \vocab{reverse mapped}). 

\subsection{Contexts}
A \vocab{context} is a tuple $(L, B, R)$, where $L$ is a DNA string called the \vocab{left part}, $B$ the base, and $R$ is a DNA string called the \vocab{right part}. The string $LBR$ is the \vocab{context string} of the context $(L, B, R)$. The context distinguishes $B$ from the rest of the context string, so that when the context is found to occur in a query string, it is clear which character in the query string (i.e. the one corresponding to $B$) has been recognized.
For an element $i$ in a DNA string $x$ a context $(L, B, R)$ is called a \vocab{natural context} if $B=x_i$,  $L$ is a (possibly empty) suffix of $(x_j)_{j=1}^{i-1}$ and $R$ is a (possibly empty) prefix of $(x_j)_{j=i+1}^n$. Some example natural contexts are visible in Supplementary Figure~S1. 

\subsubsection{Context Generality}
A context $c_1 = (L_1, B_1, R_1)$ is \vocab{forward more general} than a context $c_2 = (L_2, B_2, R_2)$ if $L_1$ is a suffix of $L_2$, $B_1 = B_2$, and $R_1$ is a prefix of $R_2$. That is, if you turned the two contexts into strings with their bases marked as special characters, the more general context would be a substring of the \vocab{less general} context. Note that a context is forward more general than itself. A context $c_1$ is \vocab{reverse more general} than a context $c_2$ if $c_1$ is forward more general than the reverse complement of $c_2$, which is $c_2^* = (R_2^*, B_2^*, L_2^*)$. We define a context $c_1$ to be generically \vocab{more general} than context $c_2$ if it is either forward more general or reverse more general than $c_2$. 

\subsection{Context-Driven Mapping}
It is possible to define a mapping scheme for a query string $x$ to a reference $G$ in terms of contexts for positions in the reference. Such a mapping scheme makes use of a context assignment.

\subsubsection{Context Assignment}
A \vocab{context assignment} assigns each position in a reference a nonempty \vocab{context set}, such that all contexts in the set have the same base as the position, and no context in one position's set is more general than any context in any other position's set (Figure \ref{fig:contextSets}). This second property of context assignments is called \vocab{nonredundancy}.  

\subsubsection{Matching}
An element $i$ in a query string $x$ is said to \vocab{match} a context $c = (L, B, R)$ if the query, when partitioned into the context $((x_j)_{j=1}^{i-1}, x_i, (x_j)_{j=i+1}^n)$, is less general than $c$. Note that this encompasses both forward less general (in which case element $i$ \vocab{forward matches} the context) and reverse less general (in which case element $i$ \vocab{reverse matches} the context). When the context is in the context set of a reference position, the element \vocab{matches} the position \vocab{on} the context. 


\subsubsection{Context-Driven Mapping Schemes}
\begin{sloppypar}
A \vocab{context-driven mapping scheme} is a mapping scheme which, for query $x$ and reference $G$ with context assignment $C$, maps each element $i$ in $x$ to the unique position in $G$ which it matches under $C$, or leaves $i$ unmapped when no such position exists. An element remains unmapped when it does not match any context of a reference position, or when it matches contexts of two or more positions; in the latter case we say it \vocab{discordantly matches}, an example of which is visible in Supplementary Figure~S2.
\end{sloppypar}

Under a (nonredundant) context assignment, each position $p$ in the reference can be mapped to, because for each context $(L, B, R)$ of $p$ the context string $LBR$ matches $p$ on that context. The nonredundancy requirement ensures this matching is not discordant: no context more general than $(L, B, R)$ can be in the context set of any other position in the reference.

\subsubsection{Stability}
An \vocab{extension} of a DNA string $x$ is a DNA string that contains $x$ as a substring.  
An element $k$ in an extension $x'$ of $x$ is a \vocab{partner} of an element $i$ in $x$  if the context $((x_j)_{j=1}^{i-1}, x_i, (x_j)_{j=i+1}^n)$ is more general than $((x'_j)_{j=1}^{k-1}, x'_k, (x'_j)_{j=k+1})$.

A mapping scheme is \vocab{weakly stable} 
if for each element $i$ in each possible query string $x$, if $i$ is mapped to a position $p$ in the reference, its partners in all extensions of $x$ will map to $p$ or be unmapped.
Weak stability is desirable because it guarantees that an element in a query cannot change its mapping to a different position under extension---the mapping scheme never has to admit that it mistook one reference position for another when presented with more information. Unlike score based mapping procedures, which are generally not weakly stable, all context-driven mapping schemes are weakly stable, because for any mapped element $i$, the partners of $i$ in an extension of the query string can only either map to the same position $p$, or be discordantly matched and therefore unmapped. This is because these partners have all the natural contexts of $i$, and therefore must match on a context in the context set of $p$, but may additionally match on the context of a different position in the reference and therefore discordantly match.

A mapping scheme is \vocab{stable} 
if for each element $i$ in each possible query string $x$, if $i$ is mapped to a position $p$ in the reference, its partners in all extensions of $x$ will map to $p$. Stability is naturally a more desirable property than weak stability, as it restricts mapping to individual positions aligned with high certainty.
By the argument above, some context-driven mapping schemes are only weakly stable. A \vocab{stable context-driven mapping scheme} is equivalent to a context-driven mapping scheme that additionally makes an element of a query string unmapped if a partner element in any extension of the query would discordantly match. 

\begin{figure}
  \begin{minipage}{0.5\linewidth}
  \centering
  \begin{tabular}{rcl}
  \multicolumn{3}{c}{Contexts for position $P_1$} \\ \hline
  $(L,$ & $b,$ & $R)$ \\ 
  \texttt{TGTCGC} & \texttt{C} & \texttt{CAAGCA} \\ 
  \texttt{TG\textbf{G}CGC} & \texttt{C} & \texttt{CAAGCA} \\ 
  \texttt{TGTCGC} & \texttt{C} & \texttt{CA\textbf{C}A} \\ 
  \end{tabular}
  \end{minipage}%
  \begin{minipage}{0.5\linewidth}
  \centering
  \begin{tabular}{rcl}
  \multicolumn{3}{c}{Contexts for position $P_2$} \\ \hline
  $(L,$ & $b,$ & $R)$ \\ 
  \texttt{ACGAC} & \texttt{C} & \texttt{CCAG} \\ 
  \texttt{CGAC} & \texttt{C} & \texttt{C\textbf{T}} \\ 
  \texttt{ACGAC} & \texttt{C} & \texttt{CCA\textbf{T}G} \\
  \end{tabular}
  \end{minipage}

  \caption{Example of two nonredundant context sets. Substitutions relative to the first context in each set are in bold. If the context $(L, B, R) = \texttt{C}, \texttt{C}, \texttt{C}$ were added to either set, it would make the context assignment redundant, as it is more general than contexts that already occur in both sets.}
  \label{fig:contextSets}
\end{figure}

\subsection{The Natural Context-Driven Mapping Scheme}

In our earlier paper \citep{paten2014mapping} we discussed a number of different context assignments, including fixed $k$-mer approaches. Here we focus on a new scheme that is easy to reason about and which performed the best in our preliminary empirical tests (Supplementary Figure~S3).


The \vocab{natural context assignment} assigns to each position in the reference the subset of its natural contexts that are not natural contexts of any other position in the reference. It is trivially nonredundant.
The \vocab{natural (context-driven mapping) scheme}, which uses the natural context assignment, has an intuitive interpretation: an element $i$ of a query string is mapped to a position $p$ of the reference when all natural contexts of $i$ with context strings unique in the reference are are assigned to $p$.

\subsubsection{Overview of Algorithms}
The natural context scheme is also simple to implement. For a reference and query, a \vocab{maximum unique match (MUM)} is a maximum length substring of the query present once in the reference. 
Our definition of a MUM differs from that used by tools like MUMmer \citep{delcher1999alignment} in that it is nonsymmetric; we allow a MUM to have multiple \vocab{MUM instances} in the query, each of which is a MUM and an interval of the query corresponding to a location of the substring. For a query $x$ of length $n$ there are at most $n$ MUM instances, since two cannot start at the same place. Each MUM instance that contains a given element $i$ can be described as a natural context string of $i$: $(x_j)^{i-1}x_i(x_j)_{i+1}$. Under the natural context assignment, the context of each such MUM-derived context string matches exactly one reference position. 

Using a suffix tree with suffix links of the strings in a reference (which can be constructed in time linear in the sum of the length of the reference strings), or a related substring index data structure, it is possible to find the set of MUM instances for a query string ordered by ascending start element in $O(n)$ time. These data structures all provide two $O(1)$ operations, \vocab{extend} and \vocab{retract}, which, given a search result set for some search string, can produce the result set for a search string one character longer or shorter, respectively. Employing these operations to find all MUMs in order by ascending query start position is straightforward. Starting with the empty string, extend the search string on its right end with successive characters from the query string until such an extension would produce a search result set with no results (or until the query string is depleted). If at that point there is one result in the result set, and at least one extension has happened since the last retraction, then a MUM has been found. Next, retract a single character from the left end of the search string, and go back to extending with the remaining unused query string characters. Repeat the whole process until the query string is depleted.


Since each successful extend operation consumes a character from the query string, no more than $O(n)$ extend operations can ever be performed. Since each retract operation moves the left end of the search string to the right in the query, no more than $O(n)$ retract operations can be performed. And since each unsuccessful extend operation (which would produce an empty result set) is followed by a retract operation, no more than $O(n)$ of those can happen either. Thus the entire algorithm is $O(n)$. 

Once the MUM instances have been found, it is necessary to identify the query elements that occur in exactly one MUM and therefore can be mapped under the natural scheme. (If an element is contained in two or more MUM instances then it must be discordantly mapped, because each must define a context that matches the element to a distinct position.) Given the MUM instances ordered by ascending query start element, it can be determined for all elements if each is in one, zero or multiple MUM instances, by a single traversal of the ordered MUM instances taking $O(n)$ time. We can therefore determine in $O(n)$ which elements in a query string are mapped. The combined time to map all the elements in a new query string given an existing reference substring index data structure of the type discussed is therefore $O(n)$.

\subsection{The $\alpha$-$\beta$-Natural Context-Driven Mapping Scheme}

Under the natural context assignment, for each (by definition minimally unique) reference context string, there must exist another reference substring that is an edit distance of one from it.  Therefore, while the natural context assignment ensures each context identifies a single position in the reference, a single substitution, insertion or deletion in a query substring could result in a change in mapping. To avoid this, we now define a more robust scheme.

Throughout, we use the Levenshtein edit distance, in which a single character replacement, insertion, or deletion is worth one. This choice of edit distance metric makes reasoning about the behavior of our algorithms simpler, but they could potentially be extended to other metrics tailored to different sequence data sources.



For a pair of overlapping substrings $(x_j)$, $(x_k)$ of a string $x$, we call elements in either substring not contained within the intersection of their intervals on $x$ \vocab{separate}.  For two substrings within the reference (not necessarily overlapping or even in the same reference string) we can similarly talk about their number of separate elements. For a given reference substring, the \vocab{$\alpha$-separation} is the minimum edit distance $\alpha$ between it and any other substring in the reference with a number of separate elements greater than its edit distance from the original substring. For a given natural context of a reference position, its $\alpha$-separation is the $\alpha$-separation of its context string.

Having a minimum $\alpha$-separation for contexts in a natural context scheme makes mappings more stable in the face of edits to the query. Specifically, it ensures that the number of edits required to transform the context of one position into the context of another is at least $\alpha$, for positions whose context strings have more than $\alpha$ separate elements. 
When two reference substrings with $\alpha$ edit distance have exactly $\alpha$ separate elements (it is easy to verify they can not have fewer than $\alpha$) then there exists a minimum edit-distance alignment of the two substrings that only aligns together bases from each substring with the same reference positions, and the $\alpha$ edit distance is therefore trivially realizable as the removal of a prefix from one substring and a suffix from the other. However, it is also possible that two substrings with $\alpha$ edit distance and $\alpha$ separate elements could have other minimum edit distance alignments that would result in different mappings. Therefore, enforcing $\alpha$-separation on a context assignment does not absolutely prevent mismappings produced by fewer than $\alpha$ edits---however, such mismappings would have to be relatively localized on the reference. 


Similarly, the natural context assignment is intolerant to edits between the query string and context strings of positions in the reference to which we might like query elements to map. To mitigate this issue, for a given context $(L,B,R)$ and element $i$ in DNA sequence $x$ we define \vocab{$\beta$-tolerance}: if $x_i = B$, $\beta$ is the minimum edit distance between the context string $LBR$ and a natural context string of $i$. If $x_i \not= B$ then $\beta=\infty$. 
Hence for a position in the reference $p$, a $\beta$-tolerant context $(L, B, R)$ is a context such that $b(p) = B$ and $LBR$ is within $\beta$ edits from a natural context string of $p$. 
The \vocab{$\alpha$-$\beta$-natural context assignment} assigns each position in the reference a context set containing the minimal length contexts that are at least $\alpha$-separated, and at most $\beta$-tolerant from it. 
It can be verified that as long as $\alpha$ is greater than or equal to one and $\beta$ is less than $\alpha/2$ then the context assignment is nonredundant and therefore valid.
The $\alpha$-$\beta$-natural context assignment ensures all admitted contexts are both $\alpha$-separated (and therefore unlikely to be matched by chance, requiring $\alpha$ misleading edits to coincide), and at most $\beta$-tolerant (and therefore tolerant of up to $\beta$ edits obscuring the true reference position). The natural context-driven mapping scheme is a special case of the \vocab{$\alpha$-$\beta$-natural (context-driven mapping) scheme} when both $\alpha$ and $\beta$ equal 0. (A possible extension would be a context scheme in which the $\alpha$-separation and $\beta$-tolerance required to admit contexts depended on the context length, but this would make the parametrization of the context scheme quite complex, and so is not explored here.)

\subsubsection{Overview of Heuristic Algorithms for the $\alpha$-$\beta$-Natural Context-Driven Mapping Scheme}

Unfortunately, algorithms built on efficient substring indexes to implement the $\alpha$-$\beta$-natural scheme require tracking a number of potential matches that is exponential in both $\alpha$ and $\beta$ parameters. Instead we pursue an algorithm that heuristically approximates this scheme. A full description of this algorithm is available in Supplementary Section~S1; the basic idea, inspired by existing seed-and-extend hashtable methods and chaining methods like BWA-MEM, is to chain exact matches separated by mismatching gaps, until a sufficient $\alpha$-separation is obtained \citep{li2010survey, li2013aligning}.

For a reference, a \vocab{minimal unique substring (MUS)} is a shortest length substring that appears once in that reference. 
Two MUSes are disjoint if they do not overlap. We define $\alpha'$ as the maximum number of disjoint MUSes within a context string. It is easy to verify that $\alpha'$ is a lower bound on $\alpha$. Intuitively, each disjoint MUS would need to be disrupted by a distinct edit.

\begin{figure}
\centering
  \includegraphics[width=1.0\columnwidth]{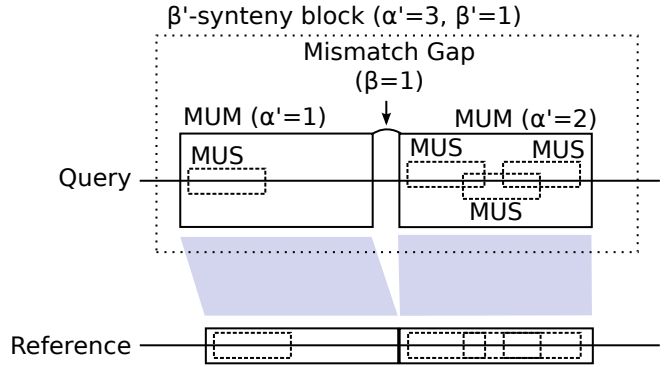}
  \caption{Diagram of a $\beta^\prime$-synteny block for the $\alpha^\prime$-$\beta^\prime$-natural context scheme, composed of two MUMs.}
  \label{fig:model}
\end{figure}

The heuristic algorithm attempts to chain together MUMs to accumulate at least $\alpha'$ disjoint MUSes, without requiring more than $\beta'$ edits in the \vocab{interstitial substrings} between the MUMs. This creates \vocab{$\beta'$-synteny blocks}, as depicted in Figure~\ref{fig:model}, which are maximal sequences of MUMs that agree in order and orientation, and which have $\beta'$ or fewer edits between the strings they mark out in the reference and the query. If a $\beta'$-synteny block can be created that has at least $\alpha'$ disjoint MUMs (and is thus $\alpha'$-separated), the MUM instances it contains are used as in the natural mapping scheme above, to define contexts for the involved reference positions.

This heuristic algorithm, as demonstrated in Supplementary Section~S1, finds contexts of reference positions in the query string that are at least $\alpha'$-separated, and at most $\beta'$-tolerant, and takes $O(\beta'^2n)$ time to map the query string, given the previously described substring index structure for the reference. Provided $\alpha^\prime < \beta^\prime$, this context scheme is nonredundant. The contexts found (and thus the matchings made) by this heuristic scheme are a subset of those that would be produced by the exact algorithm, although the same is not always true of the resulting mappings. A more thorough, empirical comparison of this heuristic scheme to an implementation of the exact scheme is left as future work, primarily due to the above-mentioned computational difficulty inherent in nontrivial exact $\beta$ values.

\subsection{Credit}

It is common to find some elements in a query string $x$ which are unmapped, and cannot be mapped on any extension of $x$, yet are intuitively recognizable as corresponding to reference positions. This often happens if bases are between two MUMs but are not part of any MUM themselves, or if they were part of a MUM between two other MUMs that cannot join a sufficiently $\alpha'$-separated $\beta'$-synteny block. In these cases, to create a scheme that maps more elements of the query,  we can augment our context assignments with additional contexts that allow such bases to map \vocab{on credit}, based on the mappings of the bases on either side. The particular credit method used here looks at the nearest mapped base on either side of a gap in mapping, and matches up elements with the correct bases with respect to their implied positions in the reference, allowing at most one mismatch. Previously unmapped elements that are matched to exactly one reference position will be mapped on credit, while elements that are matched to zero or two positions will not map.

Since only elements that did not already match, and which could not possibly match on any extension of the query, are mapped in this way, the addition of credit does not interfere with the nonredundancy of a context assignment or the stability of a context-driven mapping scheme.


\section{Results}
\label{sec:results}


In order to test the utility of the theoretical constructs described here, a series of software tests were created in order to evaluate the mappings produced by the $\alpha$-$\beta$-natural scheme described above. Mapping accuracy was evaluated for both error-corrected long sequences and error-prone short sequences.

\subsection{Mapping MHC Alt Loci}

To evaluate the performance of the new mapping algorithms proposed here a long-sequence mapping task was defined. The human genome has, on chromosome 6, a region of approximately 5 megabases known as the \vocab{major histocompatibility complex (MHC)}, rich in antigen presentation genes vital to the function of the immune system \citep{the1999complete}.
The region is prone to complex rearrangement, is well-supplied with both coding and noncoding sequence, and exhibits extreme regional variation in the polymorphism rate along its span \citep{the1999complete}. As one of the most complex and difficult regions of the genome, it provides a good testbed for methods designed to deal with difficult structural variation. To better represent the complexity of this region, the Genome Reference Consortium (GRC)'s current human reference assembly (GRCh38) contains seven full-length MHC \vocab{alt loci}, each of which serves as a different alternate reference for the region \citep{church2011modernizing}. These alt loci come with official alignments to the chromosome 6 primary sequence, which are part of GRCh38 and were generated using the NGAligner tool and limited manual curation \citep{schneider2013genome,schneider2015grc}.

\begin{figure*}[t]
  \centering
  \subfloat[]{
  	\label{fig:mhcprecisionrecall}
    \includegraphics[width=0.5\textwidth]{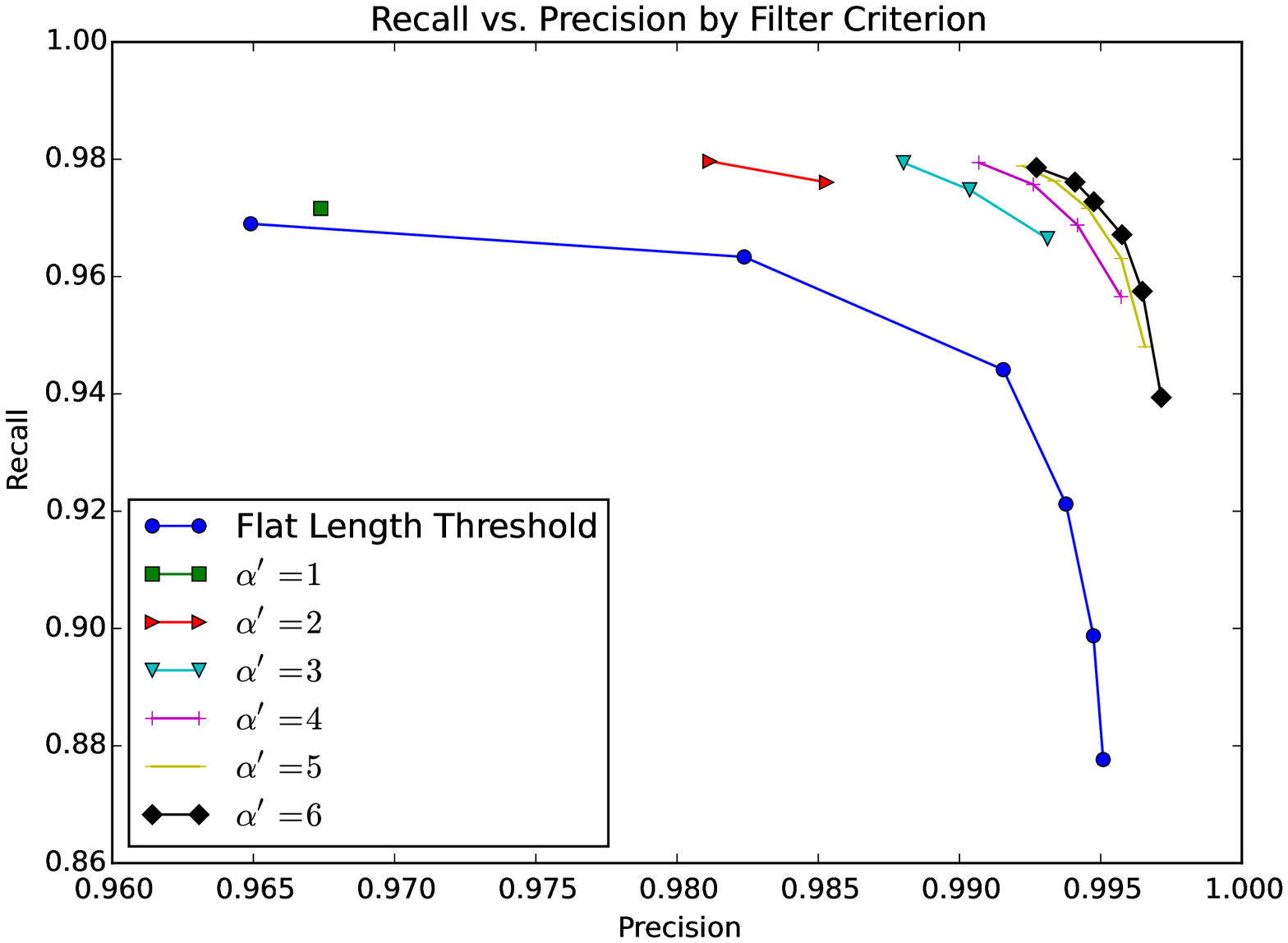}
  }
  \subfloat[]{
  	\label{fig:mhccoverage}
    \includegraphics[width=0.5\textwidth]{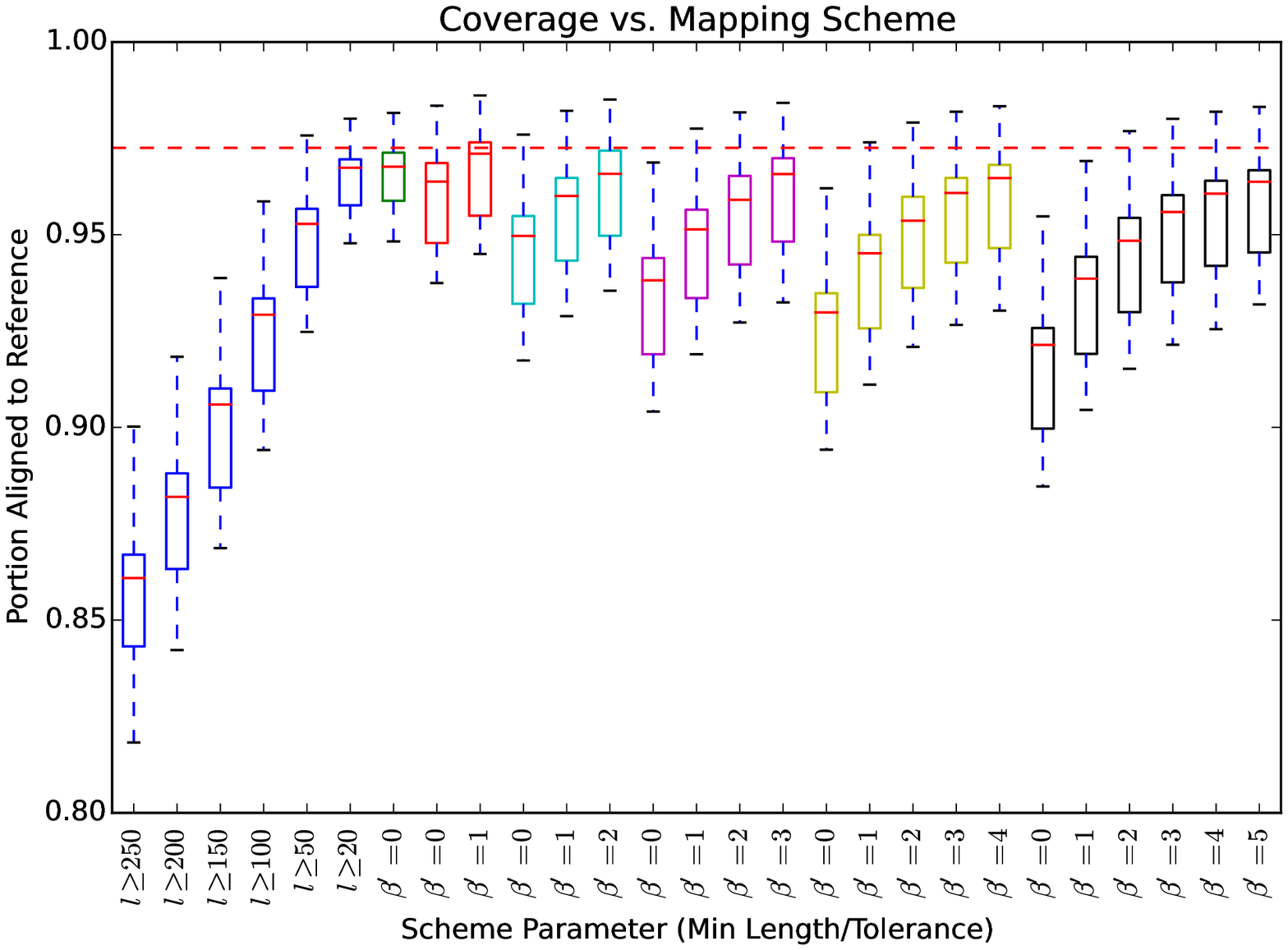}
  }
  \caption{Results of MHC alignment. Points shown in \ref{fig:mhcprecisionrecall} are averages of alt-loci alignments. Lines connect different $\beta^\prime$ levels for a given $\alpha^\prime$. The red dashed line in \ref{fig:mhccoverage} is the average coverage of the GRC reference alignments.}
  \label{fig:mhc}
\end{figure*}

\begin{figure}[t]
	\centering
    \includegraphics[width=0.3\textwidth]{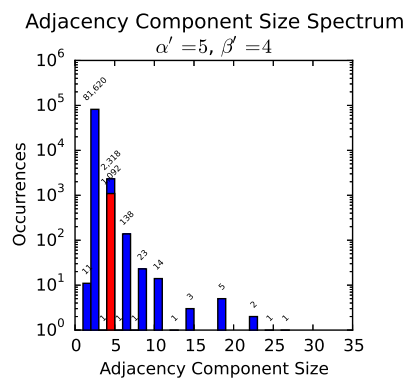}
  \caption{Frequency of rearrangements of different levels of complexity implied by the alignment of MHC alt loci to the primary path, under the $\alpha^\prime = 5$, $\beta^\prime = 4$ natural scheme, which was selected to give the best balance between precision and recall. The X axis shows the number of nodes involved in the rearrangement, while the Y axis shows the number of rearrangements of that size. The red bar shows the number of 4-node rearrangements that are automatically identifiable as tandem duplications.}
  \label{fig:rearrangements}
\end{figure}

Each mapping scheme under test took the MHC alt loci (GI568335879, GI568335954, GI568335976, GI568335986, GI568335989, GI568335992, and GI568335994), and mapped each to the GRCh38 \vocab{primary path} region, which actually appears in the ``chr6'' FASTA record. The resulting alignments were compared against the official GRC alignments distributed with GRCh38, with the caveat that aligned, mismatched bases in the GRC alignments were de-aligned to ensure a fair comparison, as the mapping schemes being evaluated were not permitted to align mismatching bases together. (Allowing mismatching bases in the GRC alignments to align made no perceptible difference in any figure, and was not pursued further.) The standard information retrieval metrics of precision and recall against the GRC alignments were calculated using \texttt{mafComparator}, and can be seen in Figure~\ref{fig:mhcprecisionrecall} \citep{earl2014alignathon}. Overall coverage (the fraction of bases of an alt locus aligned to the reference), and the frequency and size of rearrangement events implied by the alignments, were also calculated, and are visible in Figure~\ref{fig:mhccoverage} and Figure~\ref{fig:rearrangements}, respectively.


Mapping schemes using a wide range of $\alpha^\prime$ and $\beta^\prime$ parameters were tried, with $\beta'$ being restricted to values less than $\alpha'$. Additionally, the natural mapping scheme ($\alpha' = 0, \beta' = 0$) was tested, with a parameter used to vary the minimum length of admissible unique substrings (i.e. defining a series of natural context scheme, each only considering unique reference substrings longer than a minimum threshold). 

The strongest performing schemes, in terms of the harmonic mean of precision and recall (F-score), had a precision greater than 0.99 and a recall of around 0.98. Coverage was also remarkably close to that of the GRC reference alignments, suggesting that the conservative nature of the schemes did not result in undue underalignment (Figure~\ref{fig:mhccoverage}). 

In all cases the natural length-thresholded context schemes performed substantially worse than the various $\alpha^\prime$/$\beta^\prime$ combinations in terms of recall of the GRC alignments at a given level of precision (Figure~\ref{fig:mhcprecisionrecall}), and in terms of coverage (Figure~\ref{fig:mhccoverage}). This suggests that $\alpha^\prime$ and $\beta^\prime$ as defined here are effective heuristics. 

Increasing $\alpha'$ for a given $\beta'$ was found to increase precision and decrease recall, but increasing $\beta'$ at a given $\alpha'$ could restore the lost recall (at the cost of precision). The $\alpha' = 5, \beta' = 4$ natural scheme was determined to strike the best balance between precision and recall, as there was a negligible increase in precision between it and the $\alpha' = 6, \beta' = 5$ scheme (Figure~\ref{fig:mhcprecisionrecall}). Both it and the $\alpha' = 3, \beta' = 2$ scheme---selected to provide a good balance between precision and recall while also optimizing for mapping shorter sequences---were chosen for the short sequence mapping tests in \ref{subsec:reads} below.

Two additional configuration options were available for the schemes under test: whether to map unmapped internal bases on credit, and whether to enforce stability over weak stability. Our tests, the results of which are visible in Supplementary Figure~S1a and Supplementary Figure~S4b, demonstrate that requiring stability had a negligible impact on recall for long sequences, while the use of credit produced a sizable gain in recall at a manageable cost in precision (note the scales of the axes in Supplementary Figure~S4b). Consequently, credit was used for all analyses, and the stability requirement was used for the MHC mapping analysis.

The $\alpha^\prime = 5$, $\beta^\prime = 4$ natural scheme, which provided the best trade-off between precision and recall, was also evaluated in terms of the number and complexity of rearrangements it invoked to relate the MHC alt loci back to the primary path. Figure~\ref{fig:rearrangements} depicts a ``spectrum'' plot of rearrangement frequency versus size, where a rearrangement is defined as a connected component in a multi-breakpoint/adjacency 
graph representing the alignment between the primary reference sequence and an alt-loci sequence \citep{medvedev2007computability,paten2011cactus}. Briefly, the nodes of the graph are the ends (sides) of aligned sets of two or more bases and the edges the adjacencies, possibly containing interstitial unaligned sequence, that connect these ends \citep{medvedev2007computability,paten2011cactus}. The spectrum plot shows that the vast majority of rearrangements involve only two nodes (which is the structure of SNPs and simple indels), and of the rearrangements involving 4 nodes, slightly under half of them are recognizable as simple tandem duplications. The tandem duplications, which frequently involve just a handful of bases, are discoverable because of the non-linear nature of context-driven mapping. The remaining, more complex rearrangements have not been identified or named. Supplementary Figure~S5 shows UCSC Genome Browser renderings of some of the rearrangements described in the spectrum plot. 

\subsection{Mapping Simulated Short Reads}
\label{subsec:reads}

Perhaps the most important current application of traditional alignment methods is mapping reads from short read sequencing. To test this scenario a second mapping task was created. Each of the MHC alt loci sequences was broken into overlapping 200bp fragments at 100bp intervals. The read length was chosen to align with that of current or expected near future sequencing technologies, and is near the low end of what the mapping schemes presented here can accommodate \citep{quail2012tale}. Each of these fragments had substitution errors introduced with an independent probability of $1\%$ per base (comparable to current sequencing technologies) \citep{quail2012tale}. We used this simulated scenario, rather than actual reads, because it allowed us to assess the reads' origins to easily determine if mappings were correct or aberrant.

Two variants of the $\alpha'$-$\beta'$-natural scheme ($\alpha' = 3, \beta' = 2$, and $\alpha' = 5, \beta' = 4$), in both stable and weakly stable versions, were used to map each read to the primary path MHC from GRCh38. The results of the popular aligners BWA (using default parameters) and LASTZ (using an empirically-determined restrictive score cut-off), were also included \citep{li2010fast,harris2007improved}. BWA in particular functioned as a gold standard: we did not expect to outperform BWA, but rather sought to recapitulate some of its results in a context-driven framework.

\begin{figure}[t]
  \centering
  \subfloat[]{
  	\label{fig:readmappability}
    \includegraphics[width=0.5\columnwidth]{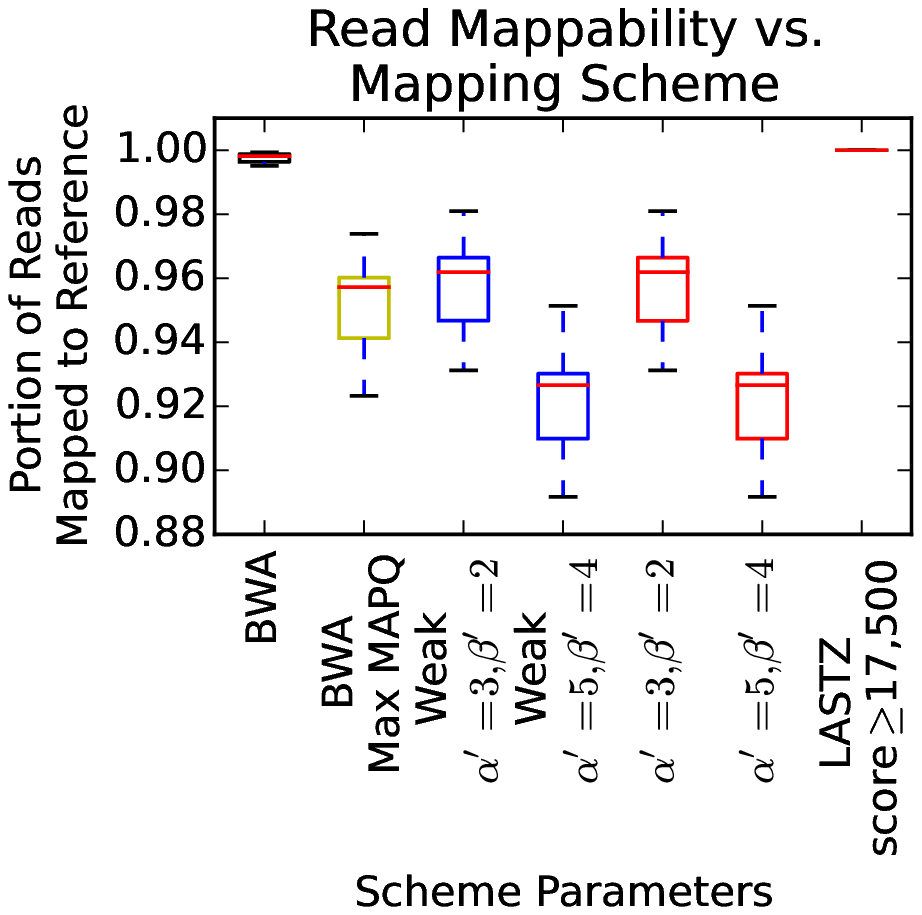}
  }
  \subfloat[]{
  	\label{fig:readcoverage}
    \includegraphics[width=0.5\columnwidth]{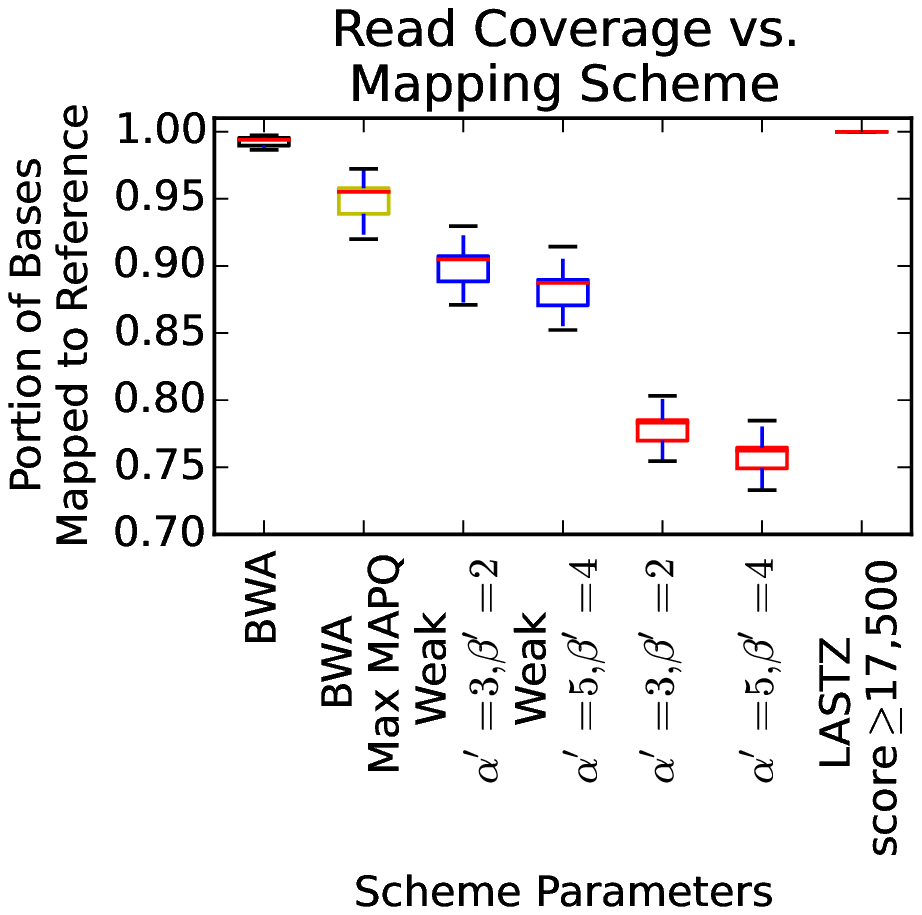}
  }
  \\
  \subfloat[]{
  	\label{fig:readgenes}
    \includegraphics[width=0.5\columnwidth]{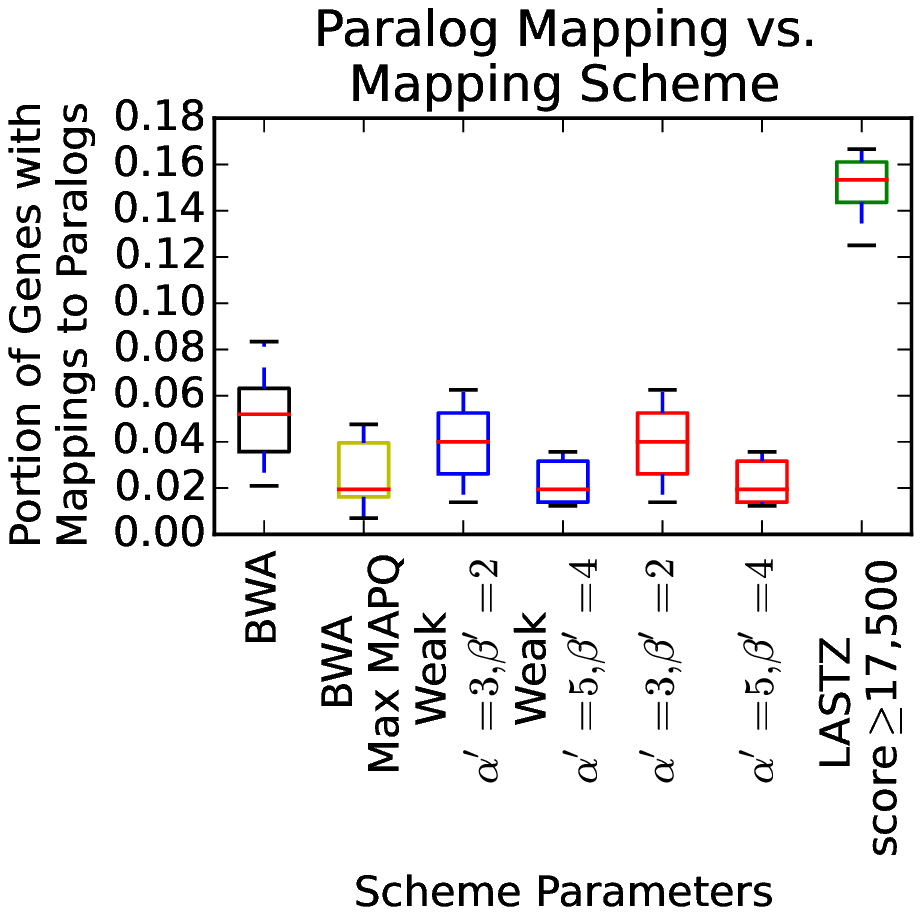}
  }
  \subfloat[]{
  	\label{fig:readwrongness}
    \includegraphics[width=0.5\columnwidth]{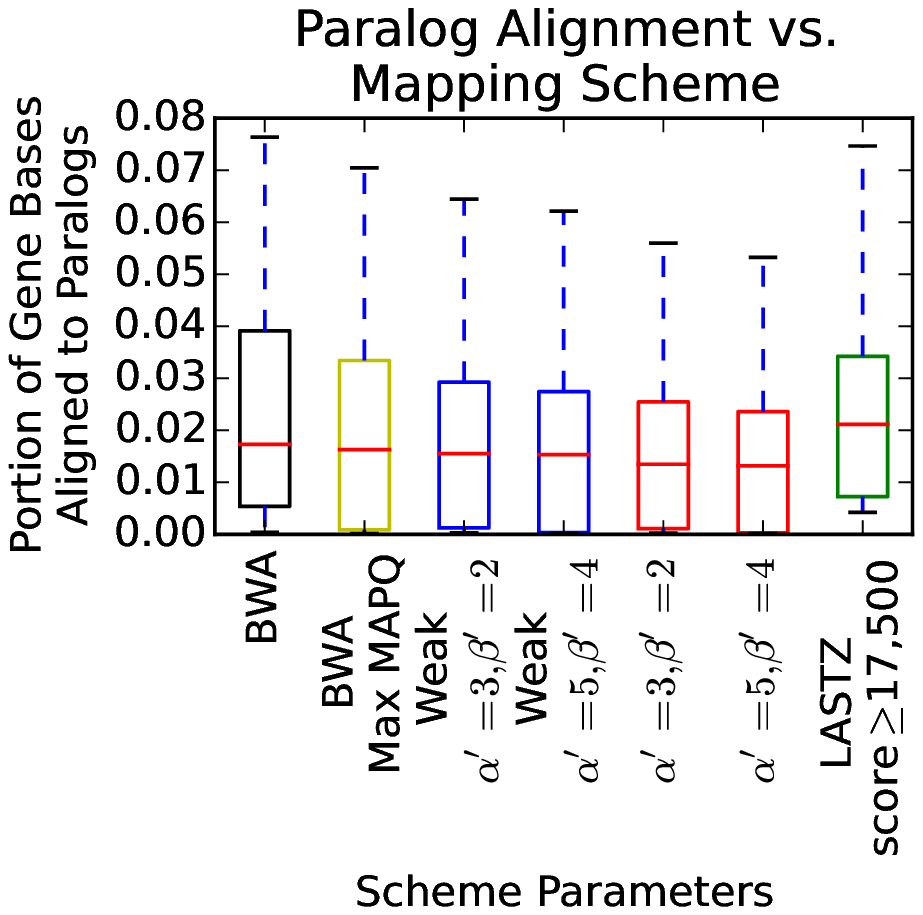}
  }
  \caption{Results of read alignments. Reads were generated from MHC alt loci by taking tiling 200bp windows at 100bp intervals, and randomly introducing substitution errors at a frequency of $1\%$. Reads were aligned to the GRCh38 primary path MHC region.}
  \label{fig:read}
\end{figure}

Mapping accuracy was assessed in two ways. First, the number of reads that each mapper could place anywhere in the reference, and portion of bases mapped regardless of correctness, were measured. These results are visible in Figures~\ref{fig:readmappability} and~\ref{fig:readcoverage}, respectively. Second, the number of genes and portion of gene bases with incorrect mappings to other genes, as annotated by the UCSC Genome Browser's ``Known Genes'' database, were also measured, and are visible in Figures~\ref{fig:readgenes} and~\ref{fig:readwrongness} \citep{meyer2013ucsc}.  


BWA and LASTZ both mapped more of the reads and covered more of the read bases than the context-driven mapping schemes, though the difference was relatively small: less than 10\% in terms of mapped reads and, for the weakly stable context schemes, less than 15\% in terms of coverage. These results were unsurprising, given that a context-driven mapping scheme is a function that can not multi-map any position, while the other two aligners freely produced multi-mappings.

The context-driven mapping schemes examined broadly matched BWA's performance in terms of avoiding mapping genes to their paralogs (Figures~\ref{fig:readgenes} and \ref{fig:readwrongness}). All four context-driven schemes tested outperformed BWA's raw output. However, if BWA's output was filtered to only include reads mapped with maximum mapping quality (which was observed to be 60), only the $\alpha'=5$, $\beta'=4$ natural schemes managed to outperform it in terms of the portion of genes with any mappings to paralogs---and that at a very substantial drop in coverage (Figure~\ref{fig:readcoverage}). LASTZ, on the other hand, did not report mapping qualities; even with what was intended to be a stringent score threshold applied, it produced the most mappings to paralogs of any aligner tested (Figures~\ref{fig:readgenes} and \ref{fig:readwrongness}). 

While the difference between stable and weakly stable mapping schemes was insignificant for long-read mapping, the coverage difference for these shorter reads was much greater. Thus stability, rather than weak stability, might seem an impractical restriction for short reads, albeit one that still admits the mapping of the majority of query sequence elements.

\begin{figure}[t]
  \centering
  \subfloat[]{
  	\label{fig:context}
    \includegraphics[width=0.5\columnwidth]{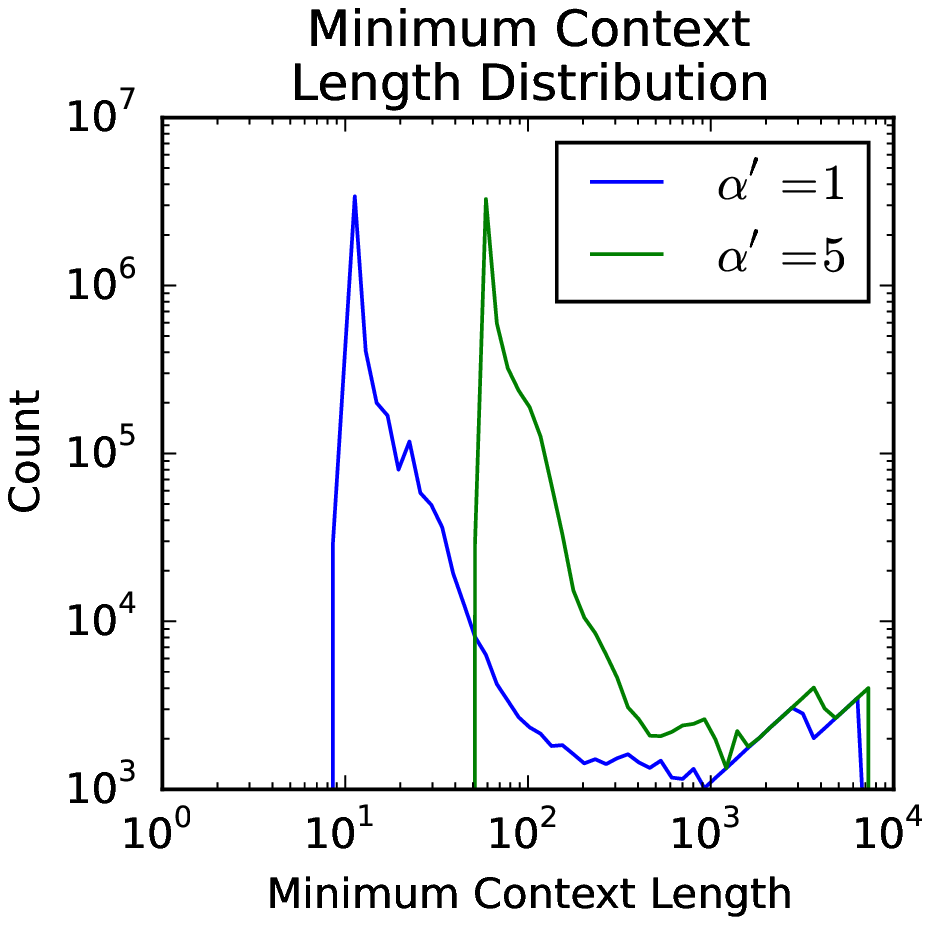}
  }
  \subfloat[]{
  	\label{fig:contextcumulative}
    \includegraphics[width=0.5\columnwidth]{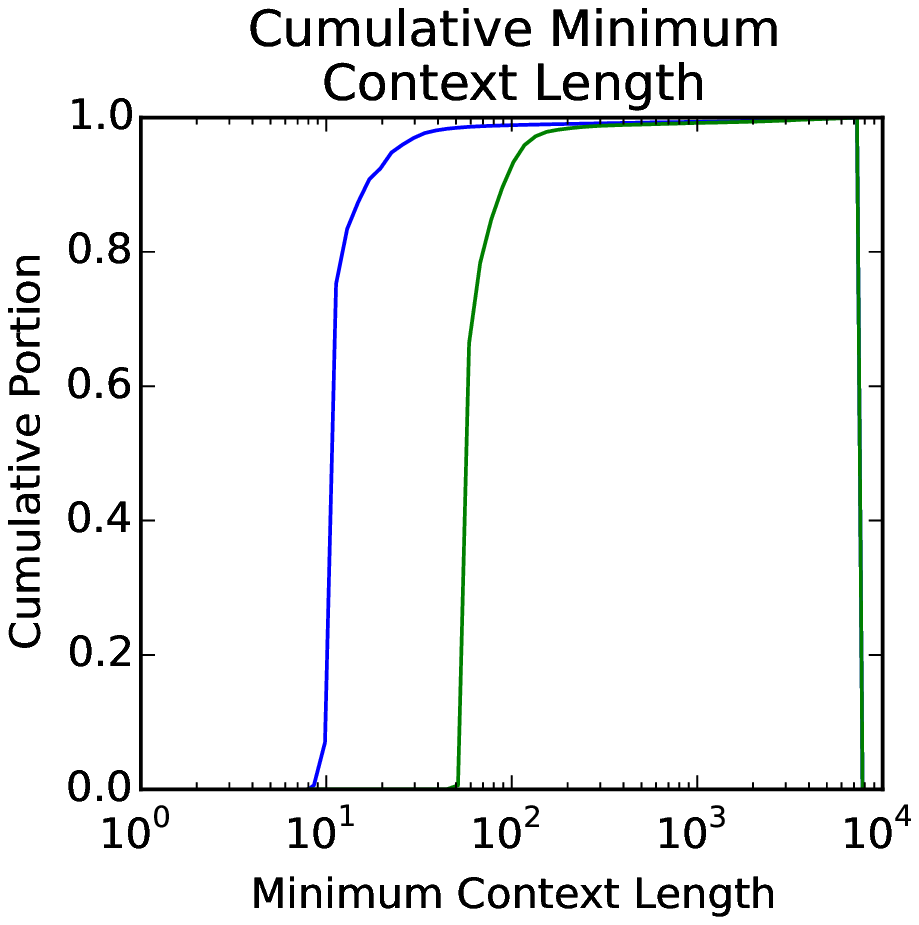}
  }
  \caption{Minimum $\beta^\prime = 0$ context lengths required to map uniquely in a reference derived from the GRCh38 primary path MHC, for different $\alpha^\prime$ values. At an $\alpha^\prime$ of $1$, $1.16\%$ of minimal contexts are longer than 100bp, and $0.97\%$ are longer than 200bp. At an $\alpha^\prime$ of $5$, $8.85\%$ of minimal contexts are longer than 100bp, and $1.74\%$ are longer than 200bp}
  \label{fig:contexts}
\end{figure}

A final experiment characterized the minimum context lengths with which it was possible for a base to map in the GRCh38 primary path MHC; the results are shown in Figure \ref{fig:contexts}. The vast majority of bases were found to be mappable with contexts of 100bp or less, and all but about $2\%$ of bases at $\alpha' = 5$ were found to be mappable with contexts of 200bp or less.

\section{Discussion}


The new mapping scheme proposed here---both radically different and more conservative than existing methods---has some important benefits. The first is that it is versatile: it can be used to map multi-megabase MHC sequences while accounting for complex rearrangements, but also does reasonably well with 200bp simulated reads. The second major benefit is stability: although requiring stability reduces coverage when mapping short reads, it reveals a majority subset of mapped positions that are aligned globally with high certainty. This is a useful per-base quality assessment somewhat orthogonal and complementary to the widely used read-level mapping quality scores \citep{li2008mapping}.
The third major benefit---being able to define variants in terms of canonical contexts which can diagnose their presence---is related to the second: having a stable mapping scheme enables the articulation of sequences which, when observed, always indicate the presence of a particular variant. This could ultimately pave the way for a high-specificity reference-free method of variant calling, building on the dbSNP concept of flanking strings \citep{sherry2001dbsnp}.



Our results show that the context-driven, stable mapping approach can be more conservative than existing mappers like BWA and LASTZ, at the cost of coverage. If there is any possibility of later having to admit that it was wrong in mapping a base, a stable scheme will not map that base. A weakly stable scheme is only slightly more permissive, willing to map bases only if it knows they cannot possibly map elsewhere. We show that the $\alpha'$-$\beta'$-natural schemes can be much more selective than LASTZ, and can in certain circumstances outperform BWA in avoiding mappings to paralogs, and in the general case are no worse. This specificity comes at the cost of a reduced ability to contend with high sequencing error rates. However, it is particularly important when analyzing regions like the MHC, where some genes present in a query may not be present themselves in the reference to which reads are being mapped, but may nonetheless have close and misleading paralogs in the reference. 



The $\alpha'$-$\beta'$-natural scheme presented here is more useful for mapping longer sequences, where the costs of stability (incurred only near the sequence ends) are lower, and the chances of finding longer and more distinctive contexts are higher. Longer reads are also more likely to directly exhibit some of the linearity-breaking structural rearrangements that our scheme is designed to deal with. The scheme presented here largely recapitulates the GRC's official alignments. The $\alpha' = 5, \beta' = 4$ instantiation, for example, has approximately $99\%$ precision and $98\%$ recall when compared to the GRC alignments, as depicted in Figure~\ref{fig:mhcprecisionrecall}. Given that the GRC alignments for the MHC alt loci do not contain any duplications, translocations, or inversions, some of the missing precision is almost certainly due to the correct detection of events that the GRC alignments did not completely describe. Judging by our manual analysis (illustrated in Supplementary Figure~S5), such calls are generally plausible. 

Finally, the context-based mapping scheme method is abstracted from its reference and query inputs, and thus easy to generalize. In addition to being very general in the types of queries it can accept, from short reads to entire alt loci, it is also very general in the types of references it can map to. As long as context sets can be defined for each position, this method can be extended to map to nonlinear, graph-based reference structures (as in Supplementary Figure~S6). Such graph structures, containing common variation in addition to the primary reference, would help to alleviate the reference allele bias inherent in current approaches to variant detection. The mapping scheme presented here provides a concrete approach to mapping to such a structure, something we explored in our earlier paper \citep{paten2014mapping} and that we are actively pursuing. 

Future work to be done on this project includes the creation of a full alignment tool based on the algorithms described here, and an extension of those algorithms to graph-structured references. The software test framework created for this work is available from \url{https://registry.hub.docker.com/u/adamnovak/sequence-graphs/}.

\section*{Acknowledgements}

\paragraph{Funding\textcolon}
This work was supported by a grant from the Simons Foundation (SFLIFE \#351901). AN was supported by research gift from Agilent Technologies, a fellowship from Edward Schulak, and an award from the ARCS Foundation. Research reported in this publication was also supported by the National Human Genome Research Institute of the National Institutes of Health under Award Number U54HG007990. The content is solely the responsibility of the authors and does not necessarily represent the official views of the National Institutes of Health.

\bibliographystyle{natbib}
\bibliography{stringmapping}

\end{document}